\documentclass[preprint,aps]{revtex4}
\usepackage{graphicx}
\usepackage{booktabs}
\usepackage[usenames, dvipsnames]{color}
\usepackage{multirow}
\usepackage{eurosym}
\usepackage{setspace}
\usepackage{amsmath}
\usepackage{mathtools}
\usepackage{graphicx}
\usepackage{epstopdf}
\usepackage{rotating}
\usepackage[ pdftex, plainpages = false, pdfpagelabels, 
                 pdfpagelayout = useoutlines,
                 bookmarks,
                 bookmarksopen = true,
                 bookmarksnumbered = true,
                 breaklinks = true,
                 linktocpage=all,
                 pagebackref=false,
                 colorlinks = true,
                 linkcolor = BrickRed,
                 urlcolor  = blue,
                 citecolor = BrickRed,
                 anchorcolor = green,
                 hyperindex = true,
                 hyperfigures
                 ]{hyperref}

\begin{document}

\title{\href{http://www.necsi.edu/blah}{Beyond Contact Tracing: Community-Based Early Detection for Ebola Response}} 
\date{\today}  
\author{Vincent Wong, Daniel Cooney and \href{http://necsi.edu/faculty/bar-yam.html}{Yaneer Bar-Yam}}
\affiliation{\href{http://www.necsi.edu}{New England Complex Systems Institute} \\ 
210 Broadway, Suite 101, Cambridge, MA 02139, USA}

\begin{abstract}
The 2014 Ebola outbreak in West Africa raised many questions about the control of infectious disease in an increasingly connected global society.  Limited availability of contact information made contact tracing difficult or impractical in combating the outbreak. We consider the development of multi-scale public health strategies and simulate policies for community-level response aimed at early screening of communities rather than individuals, as well as travel restriction to prevent community cross-contamination. Our analysis shows the policies to be effective even at a relatively low level of compliance. In our simulations, 40\% of individuals conforming to these policies is enough to stop the outbreak. Simulations with a 50\% compliance rate are consistent with the case counts in Liberia during the period of rapid decline after mid September, 2014. We also find the travel restriction to be effective at reducing the risks associated with compliance substantially below the 40\% level, shortening the outbreak and enabling efforts to be focused on affected areas. Our results suggest that the multi-scale approach can be used to further evolve public health strategy for defeating emerging epidemics.
\end{abstract}

\maketitle


\section{Introduction}

The initial medical response to Ebola in 2014 focused on caring for individuals in hospital settings and using contact tracing as the primary preventative measure  \cite{ContactTracingWHO}. Contact tracing is the accepted method for public health control of infectious diseases \cite{EbolaFAQWHO} and has been the subject of theoretical and empirical studies \cite{browne2014model, eames2003contact, tsimring2003modeling, kiss2005disease, kiss2006infectious, klinkenberg2006effectiveness, rivers2014modeling, burkeEpstein, fraser2004factors, riley2006smallpox, huerta2002contact}. Under contact tracing, a patient admitted to a hospital is asked about their recent direct contacts, and those contacts are monitored or isolated for the incubation period of the disease \cite{EbolaFAQWHO, EbolaCDC}. The spread of Ebola to dense urban communities made it difficult or impossible for contact tracing to work due to the large number of responders needed to trace contacts when many people are infected. While some contacts were knowable like family and friends, including those from ritual washing of the dead, others were untraceable anonymous interactions in public markets, buses, and taxis\cite{MonroviaRide,EbolaTransport,TaxiSpread,WHOYearReport}. For traceable interactions, the number of individuals that are needed to perform contact tracing activities (interviewing, compiling lists, seeking out contacted individuals, performing monitoring and isolation of identified individuals) grows with the number of infected cases. For an exponentially growing number of cases, the number of responders for an effective response must grow exponentially, and without this level of response, the number of cases cannot be curtailed \cite{ContactTracingWHO, CDCchallenges, Armbruster2007}. As an alternate approach, we consider community-based monitoring and limiting travel to reduce inter-community contagion \cite{YaneerDraft}. In community based monitoring, the effort to identify who has been contacted is avoided by monitoring all members of the community. Such approaches were taken in Liberia beginning in mid September 2014  \cite{BYLibSuccess, LiberiaDecrease} and in Sierra Leone beginning in mid December 2014 \cite{SierraLeoneDTD}, and may be responsible for the rapid reduction in cases seen in those two countries. In particular, in early March 2015, it was announced that there were zero remaining active cases of Ebola in Liberia \cite{LiberiaZeroNYT}. Subsequent intermittent cases have not led to the same level of crisis. While the Centers for Disease Control Director Tom Frieden attributed much of the success in the public health effort to the formation of local teams based upon the principle of RITE (Rapid Isolation and Treatment of Ebola) \cite{FriedenHuffPo}, the World Health Organization (WHO) has reported “community engagement” as a key factor in the successful response\cite{WHOYearReport}.  The direct cause of the reduction is not well documented. Here we simulate the progress of an epidemic and found that community monitoring can be highly effective in stopping an outbreak.

In general, early detection of Ebola-like symptoms is necessary for early care of patients with Ebola and limiting new infections. This is due to the extended infectious period and tendency of the disease to become more contagious as it progresses \cite{yamin2015effect, brauer2005age, towner2004rapid}. Contact tracing addresses this, but it is highly dependent on the patient knowing the people they interact with, made nearly impossible in urban environments with anonymous interactions. Here we explore a multi-scale intervention based on these ideas. Instead of monitoring individuals from a list of contacts, a community-based strategy requires that entire communities be monitored for new cases until infection is ruled out. Symptomatic individuals are isolated and treated to prevent further infections. This community-based early detection is augmented by restricting long-distance travel or subjecting travelers to extended periods of limited contact and observation. Restricting travel inhibits cross-contamination between communities and allows more targeted care to be given to infected communities. The objective is to progressively limit the disease to smaller and smaller areas and to focus resources on the areas in which the disease is present. In contrast to contact tracing, the number of responders needed for community monitoring grows only with the number of infected areas instead of the number of cases. Moreover, the effort needed to train individuals to screen for fever, the primary activity, is low, and those performing community monitoring can be members of that community. 

\begin{figure}[h!]	
\begin{center}																
\includegraphics[width=0.5\textwidth]{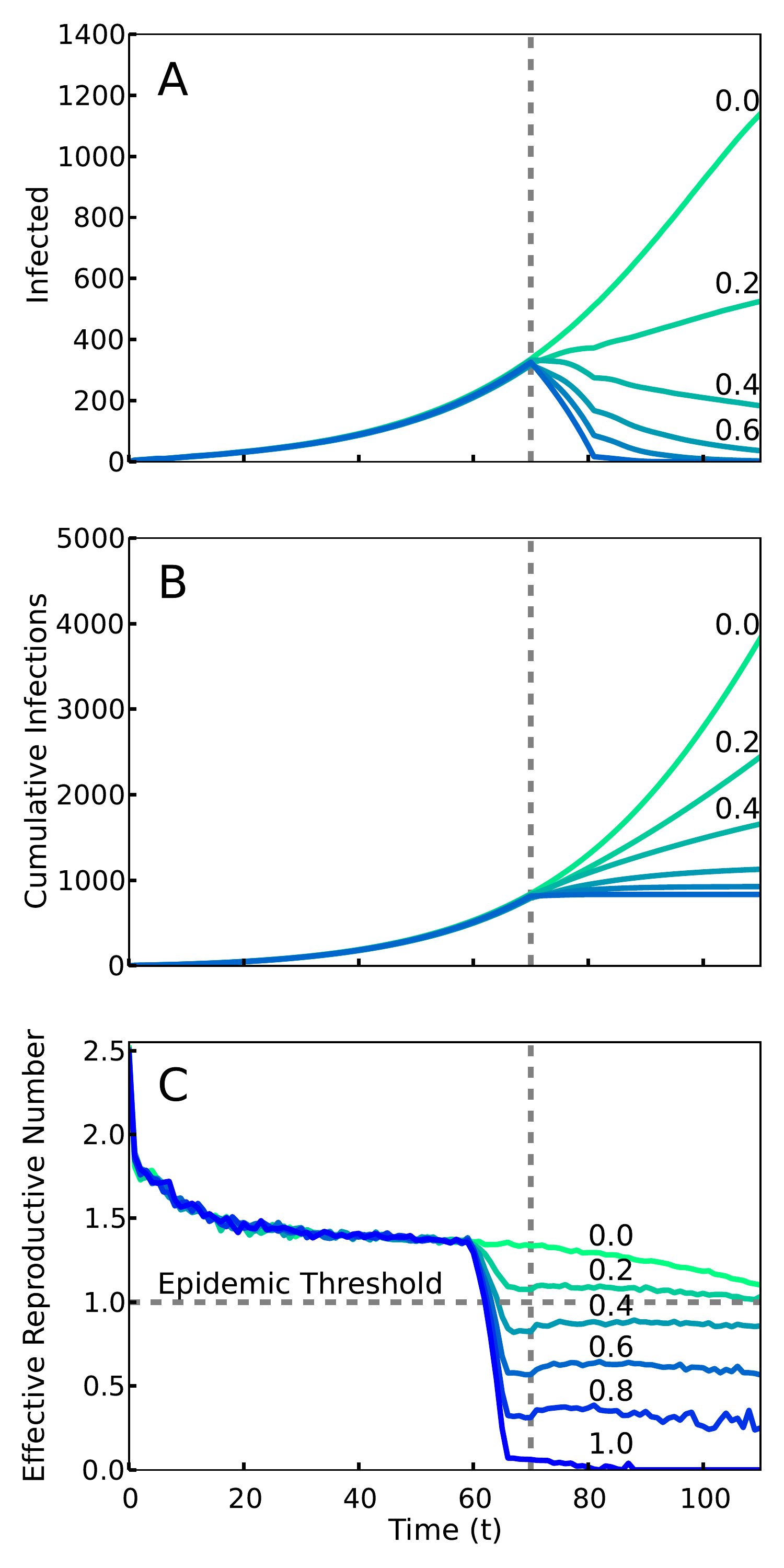}
\caption{\label{fig:1}\textbf{Simulations of an outbreak with a community-level screening intervention.} Screening begins at the vertical dotted line, with a level of compliance indicated by label and color (green $0$ to blue $1.0$). A. Number of cases with or without symptoms. Note that even 40\% compliance (0.4) results in decrease in cases. B. Cumulative cases. C. $R_t$, the effective reproductive number---the average number of individuals infected by an index case at time $t$. For an epidemic to continue to grow, $R_t$ must exceed 1. For 40\% compliance (0.4) and greater, $R_t$ decreases below one, corresponding with a decrease in active cases. $R_t$ drops before $t = 70$ because policies affect the contagion of individuals that are initially infected prior to the intervention. 
}
\end{center}
\end{figure}

If all new infections could be perfectly isolated through the complete monitoring of the population, these policies would evidently drastically limit new Ebola cases and result in an abrupt halt to the epidemic. However, an analysis must account for how effective the implementation of screening will be. It is infeasible and undesirable to constrain the population using high levels of force, so the level of compliance that is achieved is a key variable in efficacy. Here we model compliance as a probability that individuals will adhere to the community-level policies. This captures both the possibility of defiance as well as other sources of performance failure such as accidents or lack of awareness or information. We analyze the level of compliance necessary for the policies to work effectively. As shown in Fig. \ref{fig:1}, we find that even with 40\% compliance the community level policies curtail the epidemic and a 60\% compliance rapidly ends the outbreak. 

\section{Model Details}
Simulations of Ebola and other infectious diseases have been performed on complex networks \cite{keeling2005networks, eames2002modeling, pastor2001epidemic, eubank2004modelling, read2008dynamic, PhysRevE.90.012810} or spatially-explicit populations \cite{RauchBY, keeling1999effects, lopez2014addressing, fuentes1999cellular, staceygros} with interactions on household, community, and global scales \cite{PhysRevE.90.012810, kiskowski2014three, gomes2014assessing, colizza2007modeling, Colizza2008450, ball1997, frank2009}. Our analysis of response focuses on the possibility of local and long range transmission events. We incorporate such events into a transmission model and test that the results are robust to variations in the model transmission network.

\begin{figure}[h!]
\begin{center}
\includegraphics[width=0.45\textwidth]{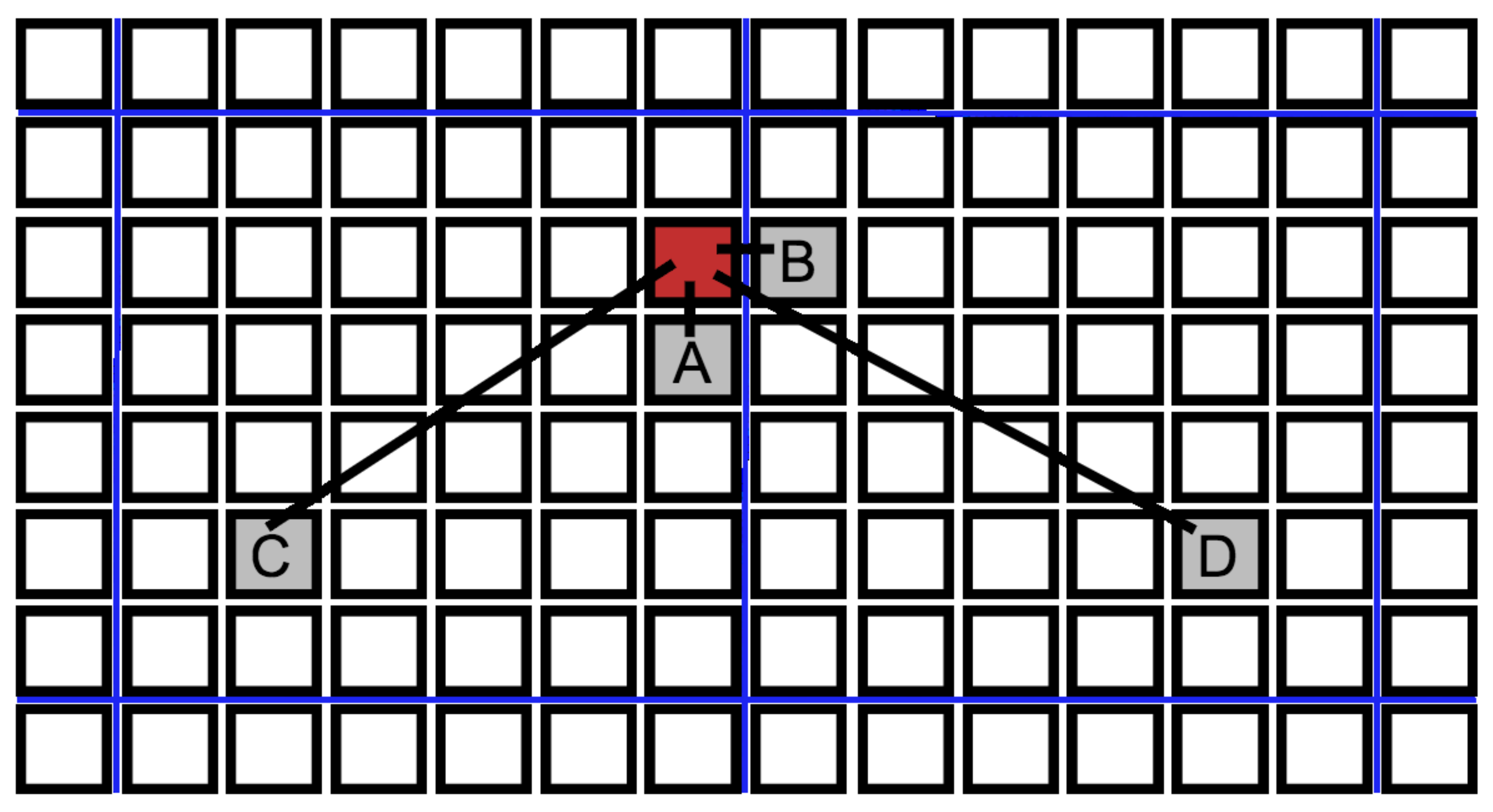}
\caption{\label{fig:2} \textbf{Schematic of different types of transmission.} Black squares indicate individuals of a spatially structured population, blue lines denote partitions between communities. A. Neighbor infection within a neighborhood. B. Cross-partition neighbor infection to another community. C. Long-range transmission within a community. D. Long-range transmission across a partition. }
\end{center}
\end{figure}

Our model is a Susceptible, Exposed, Infectious, Removed (SEIR) model on a spatial lattice of individuals (Fig. \ref{fig:2}) with periodic boundary conditions. Individuals can be in one of four states: disease-free and never previously infected (susceptible), infected in the latent period without symptoms (exposed), infected with symptoms (infectious), and recovered or dead (removed). Newly infected individuals progress through a latent period for $\Delta$ days where they are asymptomatic and not contagious. They then become contagious for a period of $\Gamma$ days, at the end of which they have either died or have recovered and have acquired immunity from further reinfection. Each individual interacts with all four of its nearest neighbors on the lattice once per day, and an infectious individual infects a susceptible neighbor with probability $\tau$ during a given interaction. Each individual also interacts with another randomly chosen individual from the population. If one of them is infectious, they have a probability $\eta$ of infecting the other by this long-range interaction. A schematic of these interactions is shown in Fig. \ref{fig:2}. This mix of local and long-range disease transmission was chosen for our model to reflect the tendency for Ebola to spread both within households and through non-local interactions in shared taxis, hospitals, or through other travel \cite{nishiura2015theoretical}. Additionally, it is known that the presence of even a small probability of long-range disease transmission can allow the rapid spread of an epidemic on a regional or global scale \cite{RauchBY}.

The community-level public health interventions are modeled based on a proposed policy draft \cite{YaneerDraft} and include daily checks for symptoms and isolation of individuals found to be in the infectious state. In our simulations, we allowed the disease to spread unabated for the first $T_{0}$ days, after which the intervention policies are put into place. The time until the start of the intervention is important in measuring the impact of early response on the control of an emerging outbreak. We assume that, even after the start of the intervention, individuals cannot be isolated and are fully capable of infecting others on the first day of their infectious period. This captures the idea that an individual can become infectious and infect someone else between symptom checks on consecutive days. With probability $\kappa$, infectious individuals are chosen to be compliant, which means that, after the first day of their infectious period, they will be perfectly isolated and incapable of interacting with others for the final $\Gamma - 1$ days of their infectious period. The lack of compliance, occurring with probability $1 - \kappa$, is assumed to be complete in the sense that noncompliant individuals continue to infect others for the duration of their infectious period.

The travel restrictions are implemented based on cordons outlined in the policy draft \cite{YaneerDraft}. 
We subdivide the population into square neighborhoods of equal size classified at each time as one of three types: $A$, $B$, or $C$. The initial type of each neighborhood is determined by monitoring of the population for $\Gamma$ days upon the onset of the public health intervention. Type $A$ neighborhoods are subpopulations in which at least one resident individual is infectious. Type $C$ neighborhoods are subpopulations for which two criteria have simultaneously been satisfied: there has not been an infectious case in the past $\Gamma$ days and the neighborhood does not share a border with any neighborhoods of type $A$. Type $B$ neighborhoods do not have any active infectious cases, but have not yet satisfied the criteria to become a type $C$ neighborhood. Neighborhoods are updated according to these criteria every timestep, however, we assume that travel restrictions implemented via a dynamic cordon prevent transmission of the disease to type $C$ neighborhoods. This ensures that residents of type $C$ neighborhoods are protected from infection, and renders type $C$ as an absorbing state. The purpose of the classification is to enable a focus of effort on screening neighborhoods of types $A$ and $B$. Given effective interventions, type $A$ neighborhoods will progressively be reclassified as type $B$ and then as type $C$, allowing for resources to be devoted to the remaining affected areas and facilitating the goal of reaching zero active cases within the region of intervention.

\section{Data and Parameters}

We used data from the World Health Organization \cite{worldebola} to determine model parameters that match the exponential growth phase of the Ebola outbreak in Liberia. By performing an exponential regression, we determined that the cumulative number of confirmed cases, $y(t)$, in Liberia approximately followed the functional form $y(t) \propto \exp{(0.052 \: t)}$, consistent with the analysis done by Chowell and Nishiura \cite{chowell2007comparative}. Comparison with empirical event data did not affect the conclusions. 

It is useful to consider the value of the basic reproduction number, $R_0$, defined as the average number of individuals infected by a single index case in an otherwise susceptible population. $R_0$ has a clear threshold value for epidemics: outbreaks with $R_0 > 1$ experience exponential growth, whereas outbreaks with $R_0 < 1$ die out exponentially. To deduce $R_0$ from the exponential growth rate, we used an expression derived from a mean-field version of the SEIR model (see Appendix):
\begin{align*} R_0 = 1 + \left(\Delta + \Gamma \right) r + \Delta \Gamma r^2  \end{align*}
where $r$ is the empirically-measured exponential growth rate of the cumulative number of cases \cite{lipsitch2003transmission}. Values for $\Delta$ and $\Gamma$ must be identified in order to obtain a value of $R_0$. We consider two different sets of parameter values corresponding to two different estimates of $R_0$ for the 2014 Ebola outbreak in Liberia. These are consistent with empirical bounds on the parameters \cite{EbolaData} and the results are robust to variation in these values as shown below.

First, we considered a latency period of $\Delta = 5$ days and an infectious period of $\Gamma = 6$ days to reflect the modeling assumptions of Althaus \cite{althaus2014estimating}. Using these time periods and infection parameters of $\tau  = 0.15$ and $\eta = 0.0125$, we conducted 1,000 simulations and obtained an average exponential growth rate of 0.054, which results in an $R_0$ estimate of 1.7, comparable to the value 1.6 obtained by Althaus. The values of $\tau$ and $\eta$ were chosen so that long range infections occurred but, as observed in West Africa, local infections dominated. The specific values do not change the conclusions. As Althaus based his parameter values on measurements from a previous outbreak on the same subtype of Ebola \cite{althaus2014estimating, team2014ebola}, the figures presented in the text of the paper are for these parameter values. 

Second, we used the parameter values $\Delta = 10$ and $\Gamma = 7$ to compare our model with that of Chowell and Nishiura, who estimated that $R_0 = 1.96$ \cite{chowell2007comparative}. Simulating this epidemic with infection parameters of $\tau = 0.18$ and $\eta = 0.015$, we obtained an average exponential growth rate of 0.05, with a corresponding $R_0$ value of 2.0, similar to that found by Chowell and Nishiura. These authors based their parameter values on previously hypothesized epidemic properties of Ebola \cite{chowell2007comparative, lekonepriors}. Results using these parameter values are in the Supplement. 

For each set of parameters, we simulated compliances ranging from 0.0 to 1.0 in steps of 0.05, and intervention delay times of 50, 70, 90, and 110. Results are averaged over 1,000 simulations of a population of 10,000 individuals, $100\times100$ square lattice, with neighborhoods of size 100, $10\times10$ sublattices, initialized with 0.02\% of the population infectious and 0.02\% latent.

\section{Results}

Fig. \ref{fig:1} shows the number of current and cumulative cases for various levels of compliance with community level interventions implemented at $T_{0} = 70$ days and without travel restrictions. We found that a relatively low compliance of 0.4 with the community screening policies was enough to end the outbreak (Fig. \ref{fig:1}A). Fig. \ref{fig:1}B shows that there is relatively little difference in the cumulative number of infections over the duration of the outbreak for the interventions with 0.6, 0.8, and 1.0 compliance. Thus, and perhaps surprisingly, the greatest gains in reducing the epidemic duration and cumulative number of cases arise from a particularly low level of compliance. 

The impact of intervention policies can be readily seen by plotting the effective reproduction number $R_t$, the average number of secondary infections caused by a primary case who is first infected at time $t$ \cite{nishiura2014early}. An $R_t$ greater than 1 at a given time implies that the epidemic will grow exponentially over the short term, while less than one implies it will decline exponentially. Unlike $R_0$, $R_t$ is designed to reflect the effect of interventions as well as that of natural epidemic burnout. Plotted in Fig. \ref{fig:1}C, we see that compliance levels of 0.6, 0.8, and 1.0 cause the value of $R_t$ to decrease well below the threshold value 1.0 within a span of several days. This represents a quick transition from a regime in which the disease is growing exponentially to a regime in which the epidemic can no longer sustain itself and consequently dies out. Even with a compliance of 0.4, enough infections were halted to reduce the value of $R_t$ below 1 (Fig. \ref{fig:1}C). 

\begin{figure*}[h!]
\begin{center}																	
\includegraphics[width=\textwidth]{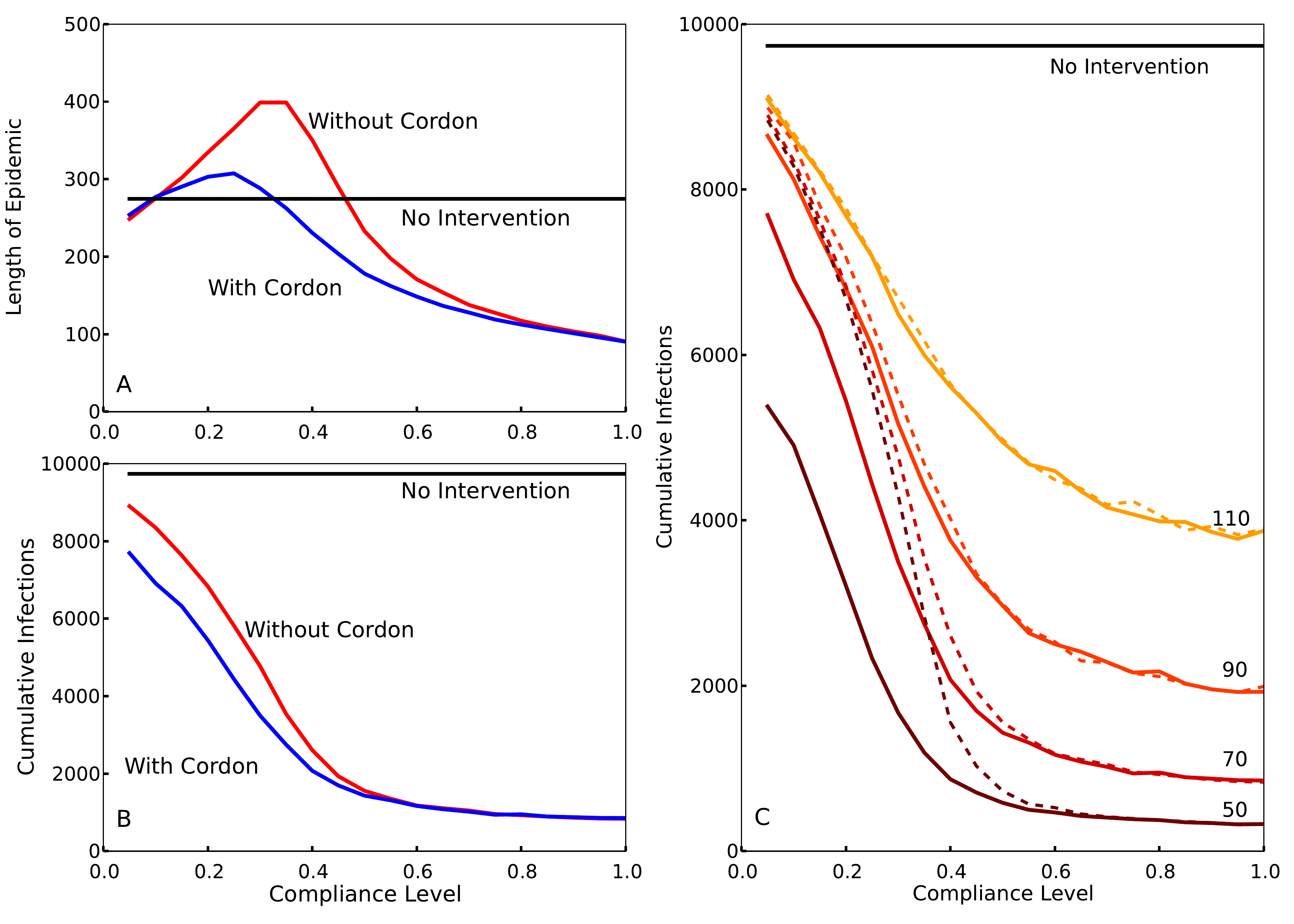}
\caption{\label{fig:3} \textbf{Effect of compliance on epidemic length and cumulative infections with and without travel restrictions.} A and B: Blue shows the case with travel restrictions, and red shows the case without such restrictions. Differentiation between the two occurs because the travel restrictions compensate for low levels of compliance. This decreases the length of the epidemic A and reduces the cumulative number of infections B in cases of low compliance. C. The cumulative number of infections over the entire epidemic, as a function of compliance levels and intervention times. Colors from brown to yellow signify intervention times (50, 70, 90, 110). Without enforced travel restrictions (dotted lines), a low compliance results in little differentiation between early and late policy implementation. The travel restrictions (solid lines) dramatically reduce infection number for earlier interventions at low compliance.
}
\end{center}
\end{figure*}

To observe the effects of the travel restrictions, we show the outbreak length and the cumulative number of infections for interventions implemented at time $T_{0} = 70$ with and without the travel restrictions for our simulated compliance values in Figs. \ref{fig:3}A and \ref{fig:3}B. Without travel restrictions, compliance levels between 20\% and 40\% actually prolong the outbreak relative to the no intervention case  (Fig. \ref{fig:3}A).  For these levels of compliance community screening slows transmission but insufficiently to halt the outbreak, so transmission continues at a slower rate until the population is exhausted. Travel restrictions greatly reduce this effect. Travel restrictions also noticeably decrease the cumulative number of infections for low levels of compliance (Fig. \ref{fig:3}B). This shows that, in the event of low compliance, travel restrictions limit the spread and duration of the outbreak. 

Comparing the cumulative cases as a function of compliance for different delay times (Fig. \ref{fig:3}C), interventions with an earlier start time $T_{0}$ generally result in fewer cumulative infections. However, without travel restrictions, this is much less true for low levels of compliance. Thus, travel restrictions ensure that early policy implementation is effective even at low compliance. Fig. \ref{fig:3}C also shows that higher compliances than 0.6 have comparatively little impact on the cumulative number of cases. 

\begin{figure*}[h!]
\begin{center}																
\includegraphics[width= 0.9\textwidth]{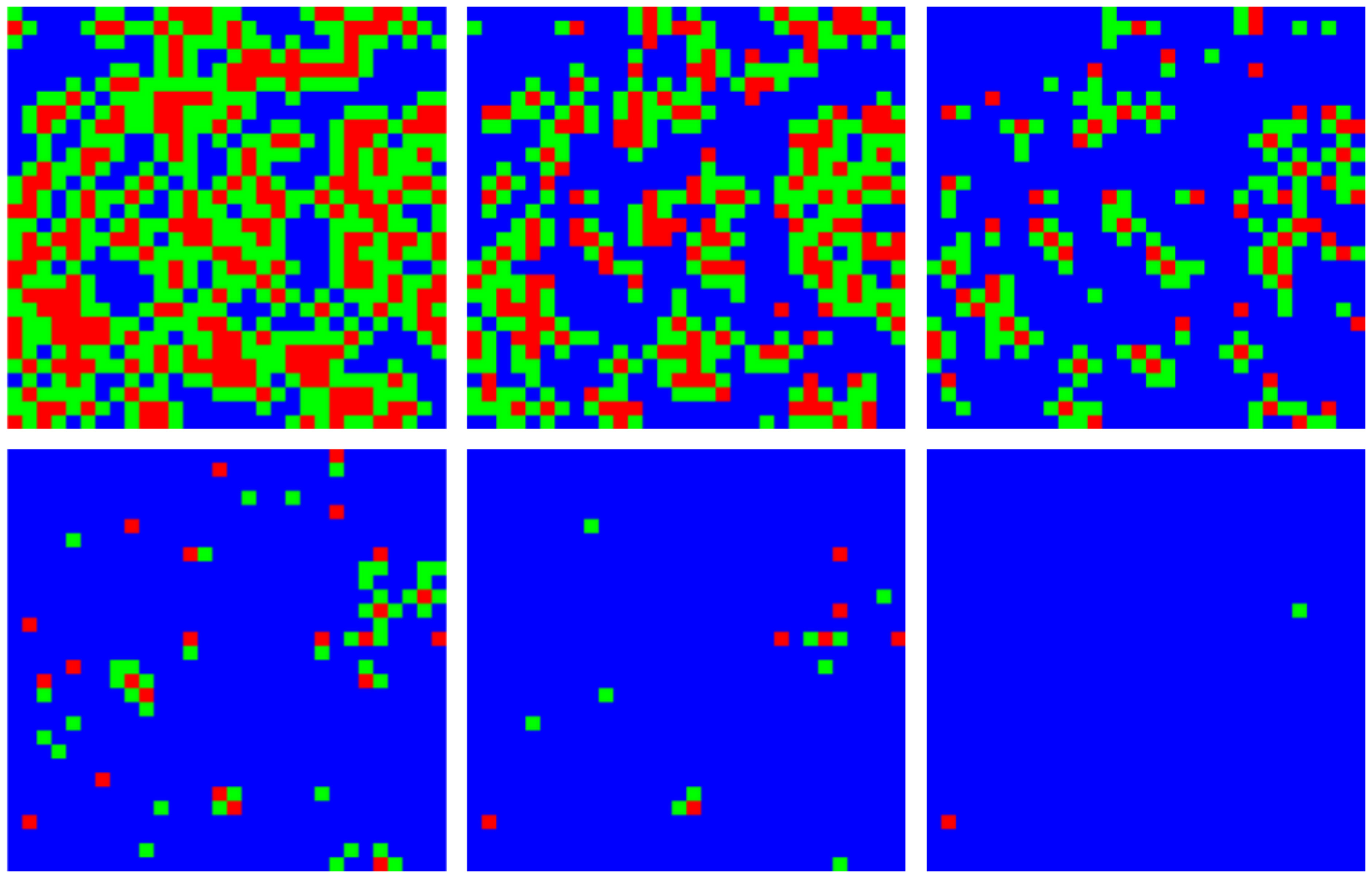} \caption{\label{fig:4} \textbf{Contraction of the epidemic areas using cordons and labeled neighborhoods.} A simulated epidemic run on a $300 \times 300$ lattice with neighborhoods of size $10 \times 10$, with 70\% compliance (0.7) and a delay of $T_{0} = 50$ days. Colored squares represent neighborhoods of types A (red, known infection), B (green, neighboring known infection), and C (blue, neither A nor B). Top (left to right) 60, 70, and 80 days, bottom 90, 100, and 110 days. Type C neighborhoods remain free from infection due to the protection provided by travel restrictions. 
}
\end{center}
\end{figure*}

We visualized the evolution of neighborhood types over time with the cordon to demonstrate the spatial constriction of the disease with this policy (Fig. \ref{fig:4}). The first panel (top left) shows the geographical distribution of neighborhood types shortly after the policies come into effect. The infected area shrinks and the ratio of type A (red) to type B (green) neighborhoods decreases over time. 

\begin{figure}[h!]	
\begin{center}																
\includegraphics[width=0.6\textwidth]{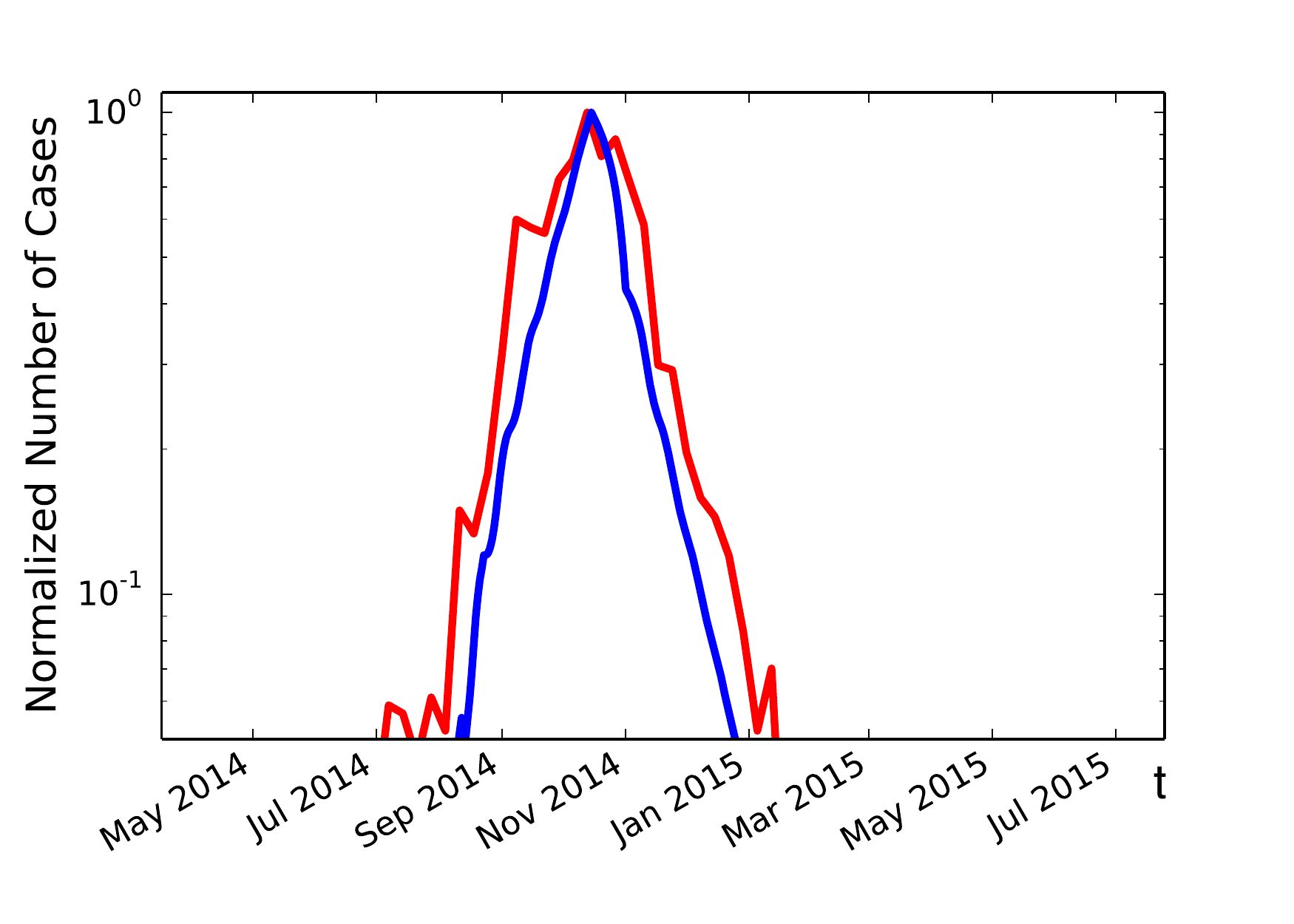}
\caption{\label{fig:6} \textbf{Comparison of empirical data with simulations} Normalized, linear-log plot of Liberia empirical values (red) compared with simulation data (blue) with $T_{0} = 50$ and 50\% compliance (0.5).
}
\end{center}
\end{figure}

In Fig. \ref{fig:6} we plot both the reported number of cases in Liberia \cite{EbolaData} and our simulation with $T_{0} = 50$ days and a 50\% compliance (0.5). Our simulations fit the observed case count data in Liberia for the parameters chosen, indicating that the early screening intervention was far from complete, but was sufficiently effective. Since the results of the simulation are robust to variation in parameters, the correspondence of the real world data in with the simulation reflects the reduction of $R_t$ below the epidemic threshold Liberia. 

\section{Robustness Analysis}

\begin{figure}[h!]	
\begin{center}
\includegraphics[width=1.0\textwidth]{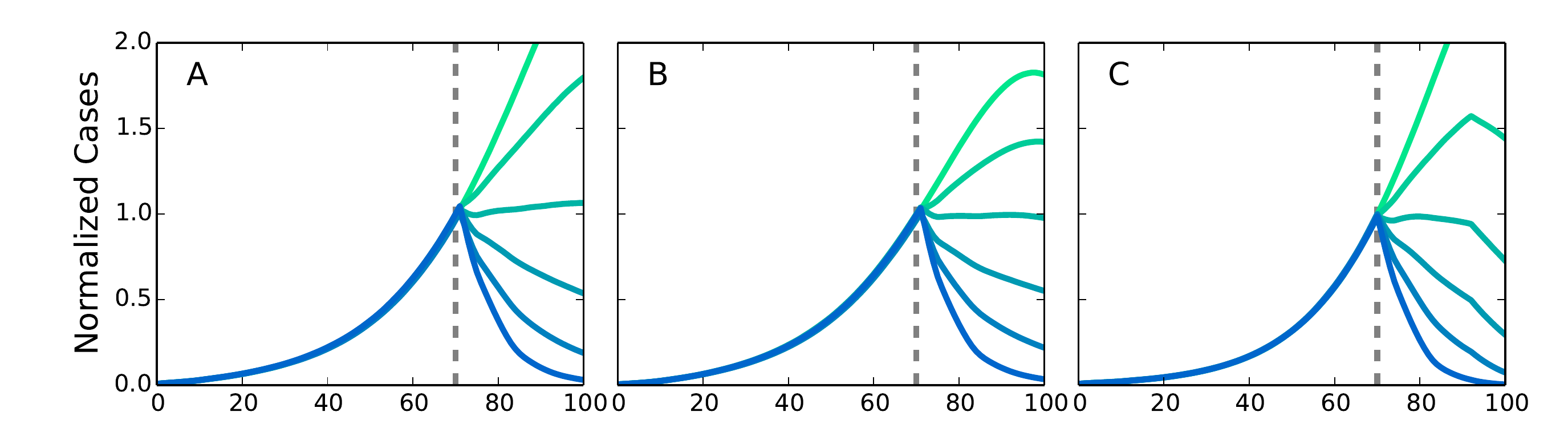}
\caption{\label{fig:5} \textbf{Simulations for different types of networks.} Screening at the dotted lines. Results are normalized to the number of infected individuals at the time of the intervention. The intervention is robust against variation in network structure. A. Von Neumann neighborhood of four nearest neighbors. B. Moore neighborhood of eight nearest neighbors. C. Kleinberg small world network, with four nearest neighbors and longrange neighbors with probability of connection decreasing as inverse distance squared. 
}
\end{center}
\end{figure}

The compliance analysis presented above is a robustness analysis of the community based response strategy. Additionally, we tested the robustness of our analysis to the approximations used to model the network of transmission. This also tests the robustness of the intervention policies to changes in the societal transmission network. We included networks in which each person had different immediate neighbors (Moore neighborhood), and ones for which the contact structure itself changes to a small world network model (Kleinberg network). For the Moore neighborhood network, each individual is connected to its closest eight rather than four nearest neighbors. The Kleinberg network was implemented by allowing individuals to have a varying number of long-range contacts with their probability of making that connection decrease with distance. We also allowed the length of the of latent and infectious periods to vary for individuals across values found by Chowell \cite{chowell2014transmission} to allow for variable stage lengths. Fig. \ref{fig:5} shows the results of the simulations with these different networks, with $\tau$ and $\eta$ adjusted so that each network's growth rate has an $R_0$ in agreement with the observed results for Ebola. The results show the intervention effectiveness remains high and a compliance above $50\%$ is sufficient to rapidly stop the epidemic. This provides additional evidence for the relevance of the results to real world conditions despite approximations used in modeling the transmission network. It also indicates that the community level interventions are highly robust. 

We note that recent research on contagious processes on networks that are not geographically local have considered heterogeneous connectivities across nodes, and specifically power law node connectivity. Under these conditions highly connected nodes enable the disease to spread across the entire network at arbitrarily low contagion rates \cite{pastor2001epidemic}. Cordons that restrict the contagion by limiting geographical spread would also impose a cutoff on power law connectivities leading to a non-zero threshold contagion rate for such networks. By limiting connectivity to local neighborhoods, cordons bound the maximum connectivity of an individual and therefore truncate power law connectivity distributions. 

We analyzed the effect of community monitoring using a mean field, spatially averaged approach. The analytic equations are constructed by considering the dynamics of infection across the entire population. The susceptible population becomes infected by short range and long range transmission from noncompliant infectious individuals. The parameters are the compliance $c$, the short range infectious rate $\tau$, the long range infection rate $\eta$, the latent period $\Delta$, the infectious period $\Gamma$, the average number of neighbors per individual $z$, the portion of neighbors that are already sick $p$, the probability that a transmission contact occurs in an unrestricted neighborhood $r$, and the probability that the transmission occurs in the same neighborhood $f$. The equations for the dynamics are: 
\begin{align*}
\frac{dS}{dt}&=-(1-c)\frac{\tau z(1-p)}{N} IS - (1-c)(1-r)\frac{\eta}{N} IS - (1-c)\frac{rf\eta}{N}IS\\
\frac{dE}{dt}&=(1-c)\frac{\tau z(1-p)}{N} IS + (1-c)(1-r)\frac{\eta}{N} IS + (1-c)\frac{rf\eta}{N}IS - \frac{1}{\Delta}E\\
\frac{dI}{dt}&=\frac{1}{\Delta}E-\frac{1}{\Gamma}I
\end{align*}
where $S$ represents susceptible, $E$ exposed, and $I$ infectious groups. The first equation can be written as
\begin{align*}
\frac{dS}{dt}&=-\frac{\tilde{\tau}}{N}IS
\end{align*}
where $\tilde{\tau}=(1-c)[\tau z(1-p)+\eta((1-r)+rf)]$ is the effective transmission rate, encapsulating all the effects of the infectiousness probabilities in the model. Note that the exposed and infected populations progress through the disease without interference. The base reproduction number without intervention $R_0$ for the SEIR model is $R_0 = \frac{\tau}{N}S \Gamma$. Therefore, the mean field value with community monitoring is $R_m = \frac{\tilde{\tau}}{N}S\Gamma$. Using this approximate mean field model, and a value of $p$ estimated from the simulation, the compliance needed to end an outbreak is $c \approx 0.5$ to reduce $R_t$ below 1.0, as compared to $c \approx 0.4$ from the simulations themselves. The consistency of the result is further indication of its robustness.

For a similar analysis of contact tracing, a mean field treatment should separate exposed and infected populations based on whether or not they were traced \cite{browne2015ebola}. Traced individuals are less likely to infect others. The success of contact tracing is dependent on the probability that an untraced infectious individual is discovered ($\rho$), and the portion of traced contacts that are isolated successfully ($\phi$). The reproduction number with contact tracing $R_c$ is $R_c = R_0(1-\rho\phi)$ \cite{browne2015ebola}. To reduce $R_c$ below 1 for ebola, $\rho\phi\geq 0.5$. This shows that the minimum of $\rho$ and $\phi$ must be more than 0.5 to reduce $R_c$ below 1 and end the outbreak. However, if one is at 0.5 they other has to be 1, i.e. 100\% successful. Otherwise, success can be achieved if both are at least 0.7. Thus both the ability to identify contacts and the successful isolation of those contacts must be sufficiently high. This makes explicit the requirement that successful contact tracing both identify individuals who have been in contact with the sick (or dead), and the need to successfully isolate those individuals.

\section{Conclusions}

We found that a policy of community response can be effective at combating disease outbreaks without relying on information about infected individual's contact networks. This highlights the possibility of alternate methods to contact tracing for combating outbreaks. We have shown the policies require a surprisingly low compliance to end the outbreak. Notably, we see that for estimated Ebola parameters, 40\% compliance is sufficient, and the cumulative number of infections in an outbreak is not substantially decreased by compliance higher than 60\%.  We also found that travel restrictions can be used to reduce the risks associated with compliance below 40\%, and that the pairing of community-level interventions and travel restrictions can result in saving a substantial fraction of individuals from infection at any level of compliance. Public health interventions implementing variants of these policies have helped the number of active Ebola cases to reach zero in Liberia in March 2015 \cite{LiberiaZeroNYT, FriedenHuffPo}. 


\begin{acknowledgments}
We thank Irving Epstein, Joseph Norman, Francisco Prieto-Castrillo, Alfredo Morales, and Matthew Hardcastle for comments on the manuscript and Joa Ja'keno Okech-Ojony, Lorenzo Dorr, Stephen Paul Ayella Ataro, Subarna Mukherjee, and Katherine Collins for help in obtaining or providing information about the Ebola response in Liberia.
\end{acknowledgments}


\newpage

\section*{APPENDIX}

\newcommand{\ddx}[2]{\frac{d #1}{d #2}}
\newcommand{\ddt}[1]{\frac{d #1}{dt}}

\subsection{SEIR Model}

We model Ebola using a Susceptible, Exposed, Infectious, Removed (SEIR) model with a finite, spatially structured population with periodic boundary conditions. In an SEIR model, each individual can be in one of four states of health:

\begin{description}
\item S (susceptible): Healthy and capable of being infected.
\item E (exposed): Infected but asymptomatic and incapable of transmitting the illness, otherwise referred to as latently infected. 
\item I (infectious): Infected and symptomatic. Capable of infecting susceptible individuals. 
\item R (removed): No longer symptomatic, infectious, or infectable. This state includes both individuals that have survived and gained immunity, and those who have died. 
\end{description}
Individuals transition from state to state in the order $S \to E \to I \to R$. Historically, the case-fatality rate for Ebola outbreaks has been around 50\% \cite{EbolaWHOFactsheet}, so the number of removed individuals can be divided by two to obtain an estimate of the number of deaths.

The standard non-spatial SEIR model is governed by a nonlinear system of differential equations. Let $S$, $E$, $I$, and $R$ represent the number of individuals in the corresponding states, then
\begin{equation} \label{eq:seir} 
\begin{array}{ll}
\ddt{S} &= - \alpha SI \vspace{2mm}  \\ 
\ddt{E} &= \alpha SI - \delta E \\ 
\ddt{I} &= \delta E - \gamma I \\ 
\ddt{R} &= \gamma I 
\end{array}
\end{equation}
where $\delta$ represents the rate at which exposed individuals become infectious and $\gamma$ represents the rate at which the disease removes infectious individuals (either via death or survival with acquired immunity) \cite{keeling2008modeling}. Transmission of the disease requires contact between individuals in state $I$ and state $S$. The infection rate parameter $\alpha$ is usually rewritten as $\frac{\kappa \tau}{N}$, where each susceptible individual has $\tau$ interactions with any of the $N$ other members of population with probability $\kappa$ of being infected by an infectious one \cite{keeling2008modeling}.

This system of differential equations provides a mean-field representation of the dynamics of an epidemic with an SEIR structure. We consider potential policy recommendations that involve explicit change of the contact network structure of the population. For our purpose, a model based upon this mass-action system of differential equations is insufficient. 

\subsection{Spatial Model}

We model a population on a square lattice where each individual has three properties: their current state of health ($S$, $E$, $I$, or $R$), the amount of time for which they have been in that state, and whether or not they are compliant with community-level policies. If an individual is characterized as compliant, then they can be successfully isolated after entering the infectious state.

Using the definitions from the SEIR model, we consider the rate of transition from state $E$ to state $I$ to be $\delta$ and the rate of transition from state $I$ to state $R$ as $\gamma$. For simplicity, individuals transition deterministically from state E to state I (from state I to state R) after $\Delta = \frac{1}{\delta}$ ( $\Gamma = \frac{1}{\gamma}$) time steps. We take $\Delta$ and $\Gamma$ to be integer numbers of days so that we can choose each time step of the simulation to represent a single day. 

We initialized the population by randomly setting 0.02\% of the individuals to be in each of states $E$ and $I$, with the remaining individuals starting out in state $S$. Each individual was designated as compliant with probability $\kappa$. All individuals seeded in state $I$ were initialized at the beginning of the infectious period, and individuals seeded in state $E$ were given a random number between $\Delta$ and $\frac{3}{5} \Delta$ days remaining in the latent period. We chose 0.02\% and the initial compartment times in order to smoothy simulate the epidemic growth. Similations with different small initial numbers of seed cases and seeded state times yielded essentially the same behavior. 

The transmission of the disease involves both local and long-range spreading mechanisms. Each infectious individual has probability $\tau$ of infecting each of its susceptible neighbors during a given day. Additionally, each susceptible individual chooses at random an individual on the grid with whom they will interact on a given day and, if infectious, the susceptible will become infected with probability $\eta$. The formal value of the network connectivity is N as all nodes are connected by the long range transmission probability.

\subsection{Spatial model $R_0$}

The basic reproduction number, $R_0$, is generally defined as the expected number of individuals that a single seeded individual in state $I$ will infect if the rest of the population is susceptible. For large population size $G$, the $R_0$ value of this process, with infectious period $\Gamma$ and number of neighbors ($z=4$), is approximately given by  
\begin{equation} \label{eq:Rderived}
R_0  \approx  \Gamma \eta  + z \left(1 - \left( 1 - \tau \right)^{\Gamma}\right)  
\end{equation}
The number of neighbors that are infected is given by the second term, and the number of non-neighbors that are infected is given by the first term. The number of neighbors that become infected is complicated by the reduction in number of susceptible neighbors as they become infected from day to day during the infectious period. For the long range interactions, the effect is small due to the large number of possible sites so that a few new infections do not affect the number of susceptible individuals the long range interactions can affect. 

 To obtain equation \ref{eq:Rderived}, consider a neighbor that can be infected by the local infection process with probability $\tau$ and independently, by the long-range infection mechanism with probability $\frac{\eta}{G^2}$. The probability that the neighbor is not infected on a given day is $\left(1 - \tau \right) \left(1 - \frac{\eta}{G^2} \right)$. An individual is infectious for $\Gamma$ days after becoming infected. This means that the probability of a particular neighbor being infected is 
\begin{equation}
1 - \left( \left(1 - \tau \right) \left(1 - \frac{\eta}{G^2} \right) \right)^{\Gamma} = 1 - \left( 1 - \tau \right)^{\Gamma} \left(1 - \frac{\eta}{G^2} \right) ^{\Gamma}
\end{equation}
For z neighbors that can be independently infected (periodic boundary conditions ensure any such individual has the same number of neighbors), the number of infected neighbors is (neglecting corrections of $O(1/G^2)$)
\begin{equation}
z \left(1  - \left( 1 - \tau \right)^{\Gamma} \left(1 - \frac{\eta}{G^2} \right) ^{\Gamma} \right) 
= z \left(1 - (1 -\tau)^{\Gamma} + O \left(\frac{1}{G^2} \right) \right)  
\end{equation}
For large $G$ the number of infected individuals in the neighborhood is thus $z \left(1 - (1 - \tau)^{\Gamma} \right)$.

Individuals outside of the infected individual's neighborhood can only be infected by long-range interaction. Neglecting corrections of $O(1/G^2)$, the infected individual chooses one such individual to visit and infects that individual according to the probability $\eta$. Since it is assumed that the concentration of infected individuals is small, the average number of individuals infected is $\eta$. The additional infection is only $1$ in $G^2$, which doesn't affect the calculation in the next period, so the total number over the infectious period is $\Gamma \eta$.

We note that the characterization of $R_0$ differs from the differential equation SEIR (mean field) model. Since an individual locally infected by the first seeded case has that seeded case as a neighbor, the expected number of susceptible neighbors is less than the seeded case. Therefore, it is unjustified to assume the calculation of $R_0$ for the first individual holds for the rest of the contagion. It is better to think of the value of $R_0$ as the contagion rate in the low density limit rather than, as is it conventionally discussed, that of a single individual. 

The dynamics of the epidemic can be more completely described by the effective reproduction number, $R_t$, defined as the average number of secondary infections caused by an index case that is infected at time $t$. This includes the effect of the reduction of the number of susceptible individuals (epidemic burnout) and susceptible neighbors \cite{RauchBY} as well as the impact of the community-level intervention at varying levels of compliance. $R_t$ can be approximated by 

\begin{equation} \label{eq:SandZ}
R_t \approx  \Gamma \eta s_t + z_t \left(1 - (1 -\tau)^{\Gamma}  \right)  
\end{equation} 
where $z_t$ is the average number of susceptible neighbors for an individual infected at time $t$ and $s_t$ is the proportion of susceptible individuals in the whole population. Eq. \ref{eq:SandZ} reduces to Eq. \ref{eq:Rderived} if a susceptible population is seeded with a single infectious individual, as $s_t \approx 1$ and $z_t = z$, which is an individual's neighborhood size in the given population structure. 

\begin{figure}[h!]
\begin{center}
\includegraphics[width=  \textwidth]{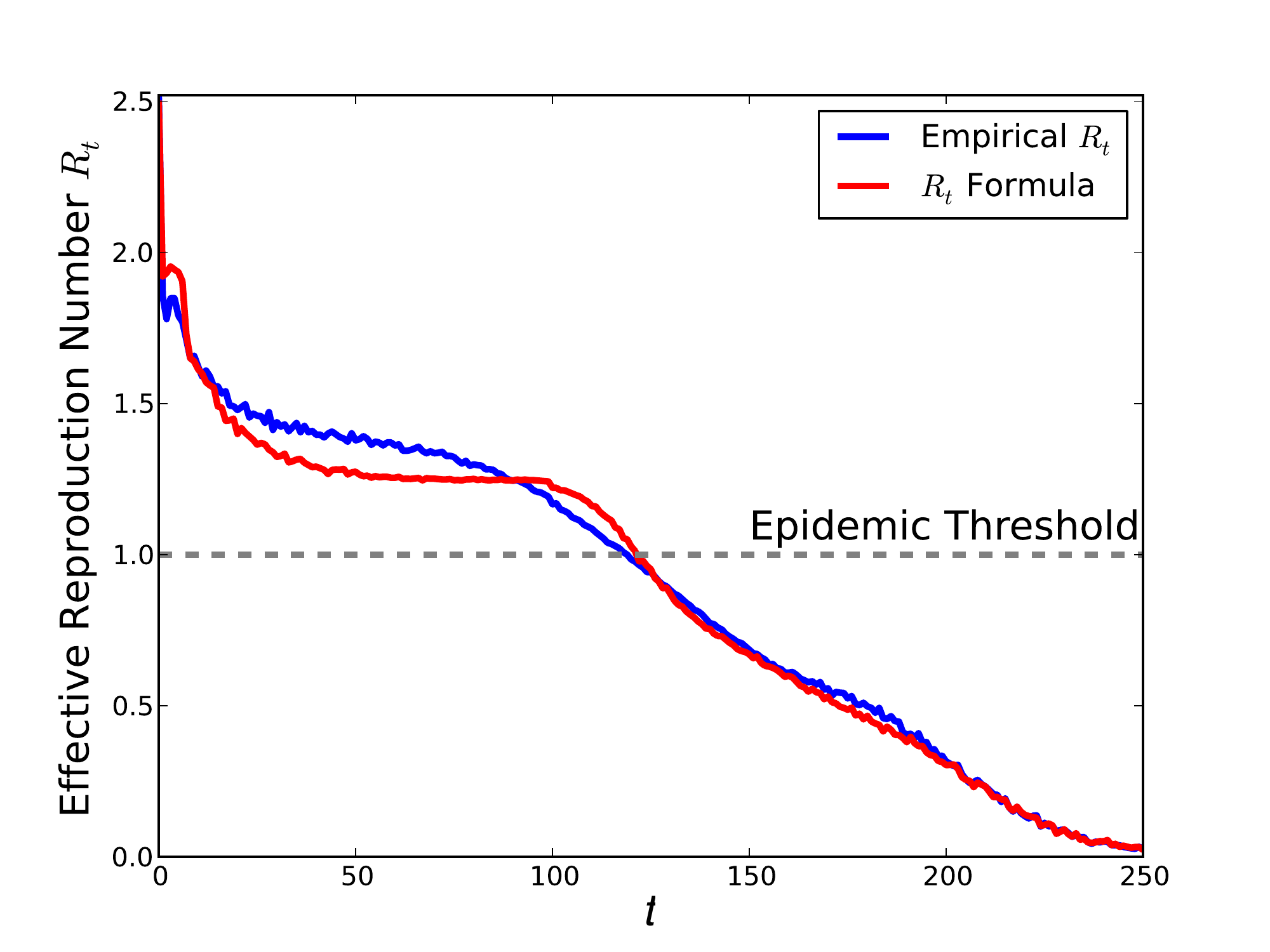}
\caption{\label{fig:rtcompare} \textbf{Comparison of $R_t$.} The value of $R_t$ measured from the average number of secondary infections caused by an individual infectied at time $t$, averaged over 1,000 simulations, is shown in blue. $R_t$ calculated by Eq. \ref{eq:SandZ} is shown in red. The values of $z_t$ and $S_t$, the average number of susceptible neighbors for an individual infected at time $t$ and the average number of susceptibles in the population at time $t$,  are also obtained from an average over 1,000 simulations. }
\end{center}
\end{figure}

In Fig. \ref{fig:rtcompare}, we compare the simulated time-series of values for $R_t$ with the time-series of $R_t$ generated by Eq. \ref{eq:SandZ} using simulated values of $z_t$ and $s_t$ averaged over 1,000 simulations with $\Delta = 5$, $\Gamma = 6$, $\tau = 0.15$, and $\eta = 0.0125$ and no public health intervention. Eq. \ref{eq:SandZ} agrees well with the values of $R_t$. The value of $R_t$ (2.52) is consistent with Eq. \ref{eq:Rderived} (2.57). 
 
\subsection{Model Parameters and Epidemiological Analysis}

 In order to make relevant comparisons between the results of our simulations and the actual 2014 Ebola outbreak, it is useful to choose model parameters so that the spread of our simulated epidemic matches the real-world spread of Ebola. We chose parameters so that the simulated epidemic matches the growth of cumulative case numbers observed in the Liberia outbreak during the period of exponential growth. Values obtained by comparison with event data and other methods do not change the conclusions. In addition, we consider values of $\Delta$ and $\Gamma$ consistent with the actual mean latent and infectious periods for Ebola \cite{EbolaData} and the results are robust to variation in these values as shown below.. Given these values of $\Delta$ and $\Gamma$, we find values of the infection parameters $\tau$ and $\eta$ so that the growth rate of cumulative cases in the simulated epidemic is consistent with the actual outbreak.  
 
 Given a time series of cumulative cases, $x(t)$, one can estimate the rate of initial exponential growth from a fit to the expression  $\ln \left(x(t)\right)= b + \lambda t$ \cite{chowell2007comparative}. Using data from the World Health Organization, we find $\lambda = 0.052$ for the 2014 outbreak in Liberia, consistent with the value computed by Chowell and Nishiura \cite{chowell2014transmission}. 
 
 We consider two sets of $\Delta$ and $\Gamma$ that have been used to describe the Ebola outbreak. First, we use $\Delta = 5$, $\Gamma = 6$, approximating values of Althaus \cite{althaus2014estimating} ($\Delta = 5.3$, $\Gamma = 5.61$) taken from a previous Ebola outbreak with similar immunological properties. We found that this $\Gamma$ and $\Delta$, when paired with infection parameters $\tau = 0.15$ and $\eta = 0.0125$, produced a Liberia-like exponential growth rate of 0.0536. The results from simulations with these parameter values can be found in the main paper. 
 
  We also considered the values $\Delta = 10$, $\Gamma = 7$, which are consistent with the values used by Chowell and Nishiura \cite{chowell2014transmission} ($\Delta = 10.1$, $\Gamma = 6.5$), who used hypothesized parameter values proposed by Lekone and Finkenstadt \cite{chowell2014transmission, lekonepriors}. Using infection parameters of $\tau = 0.18$ and $\eta = 0.015$, our simulated epidemic with this $\Delta$ and $\Gamma$ produced an exponential growth rate of 0.0505, which also roughly matches the exponential growth rate in Liberia. Results from these simulations can be found later in the appendix. 

It is common to characterize simulated and real-world outbreaks using $R_0$, the average number of secondary infectious caused by a single infectious individual in an otherwise susceptible population. One can obtain an expression for $R_0$ in terms of $\Delta$, $\Gamma$, and $\lambda$, the empirically observable exponential growth rate of cumulative cases, through a linear stability analysis of the mean field SEIR model. 

The cumulative number of cases $x(t)$ satisfies $x(t) \propto e^{\lambda t}$ near the disease-free equilibrium, where $\lambda$ is the dominant eigenvalue of the  Jacobian matrix obtained from linearizing Eq. \ref{eq:seir} around the disease free equilibrium (i.e. where $E = I = R = 0$ and $S = N$)   \cite{chowell2004basic, chowell2007comparative, lipsitch2003transmission, heffernan2005perspectives}. It can be shown that $R_0$ is equal to
\begin{equation} \label{eq:Rapproxsim} 
R_0 = 1 + \left( \Delta + \Gamma \right) \lambda + \left( \Delta \Gamma \right) \lambda^2 
\end{equation}

Using Eq. \ref{eq:Rapproxsim} and the exponential growth rates generated from our simulations, we see that $R_0$ is approximately equal to 1.68 for the $\Delta = 5$, $\Gamma = 6$ simulations, and is equal to 2.04 for the $\Delta = 10$, $\Gamma = 7$ case. These values of $R_0$ agree well with the values of $1.59$ and $1.96$ estimated by Althaus \cite{althaus2014estimating} and by Chowell and Nishiura \cite{chowell2014transmission}, respectively. This provides further confirmation that our simulated epidemics display similar exponential growth behavior to the 2014 outbreak in Liberia.

\subsection{Results for $\Delta = 10$, $\Gamma = 7$}

\begin{figure}[h!]
\begin{center}
\includegraphics[width=0.5\textwidth]{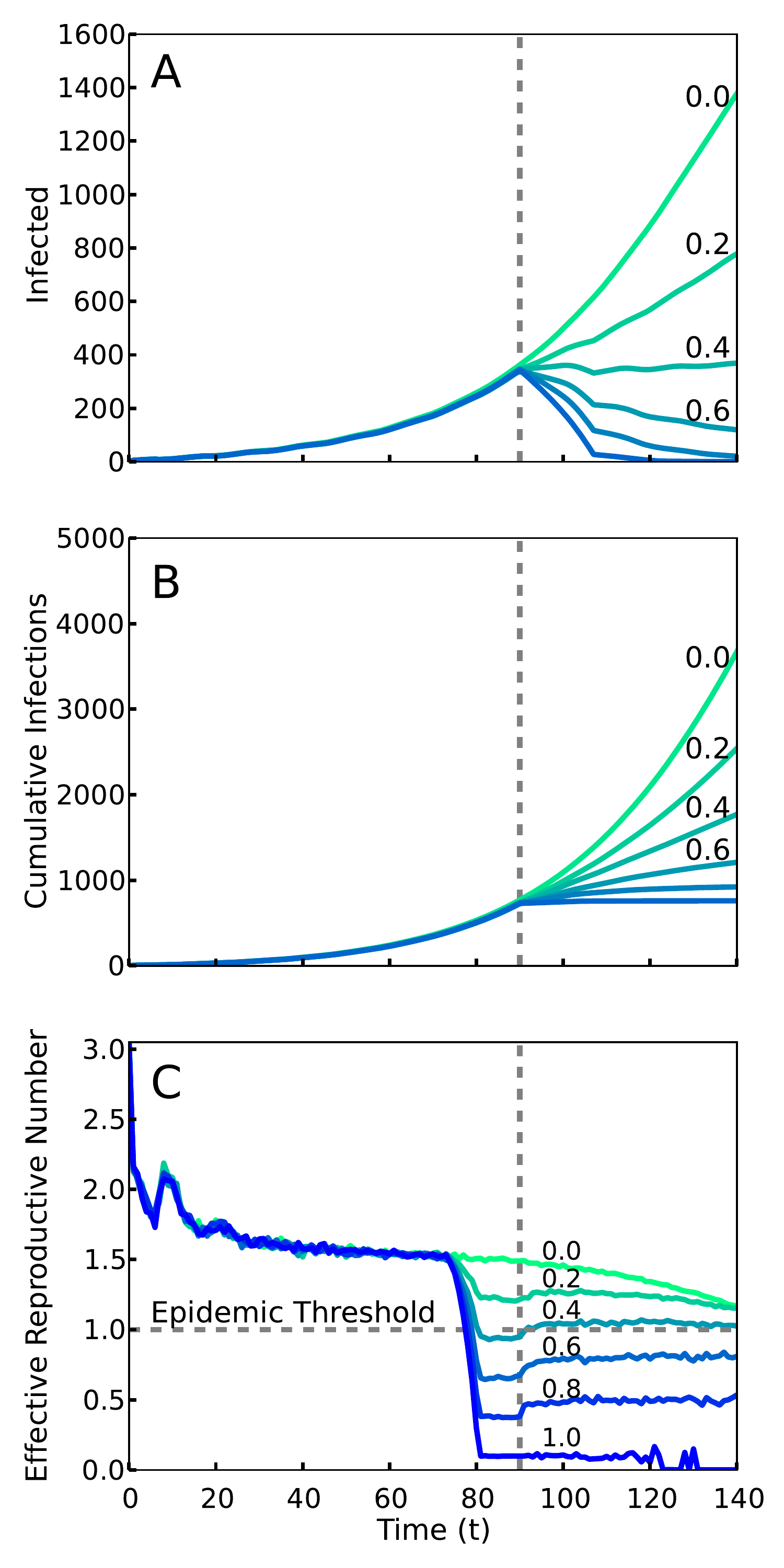}
\caption{\label{fig:1appendix}\textbf{Simulations of an outbreak with a community-level screening intervention.} Screening begins at the vertical dotted line, with a level of compliance indicated by label and color (green $0$ to blue $1.0$). A. Number of cases with or without symptoms. Note that, compared to the simulations in the main paper, 40\% compliance (0.4) is no longer sufficient to end this more virulent outbreak. B. Cumulative cases. C. For greater than 40\% compliance (0.4), $R_t$ decreases below one, corresponding to a rapid decrease in active cases. Despite this change, the overall results are robust as a compliance value of 0.6 is sufficient to end the outbreak.}
\end{center}
\end{figure}

\begin{figure}[h!]
\begin{center}
\includegraphics[width= \textwidth]{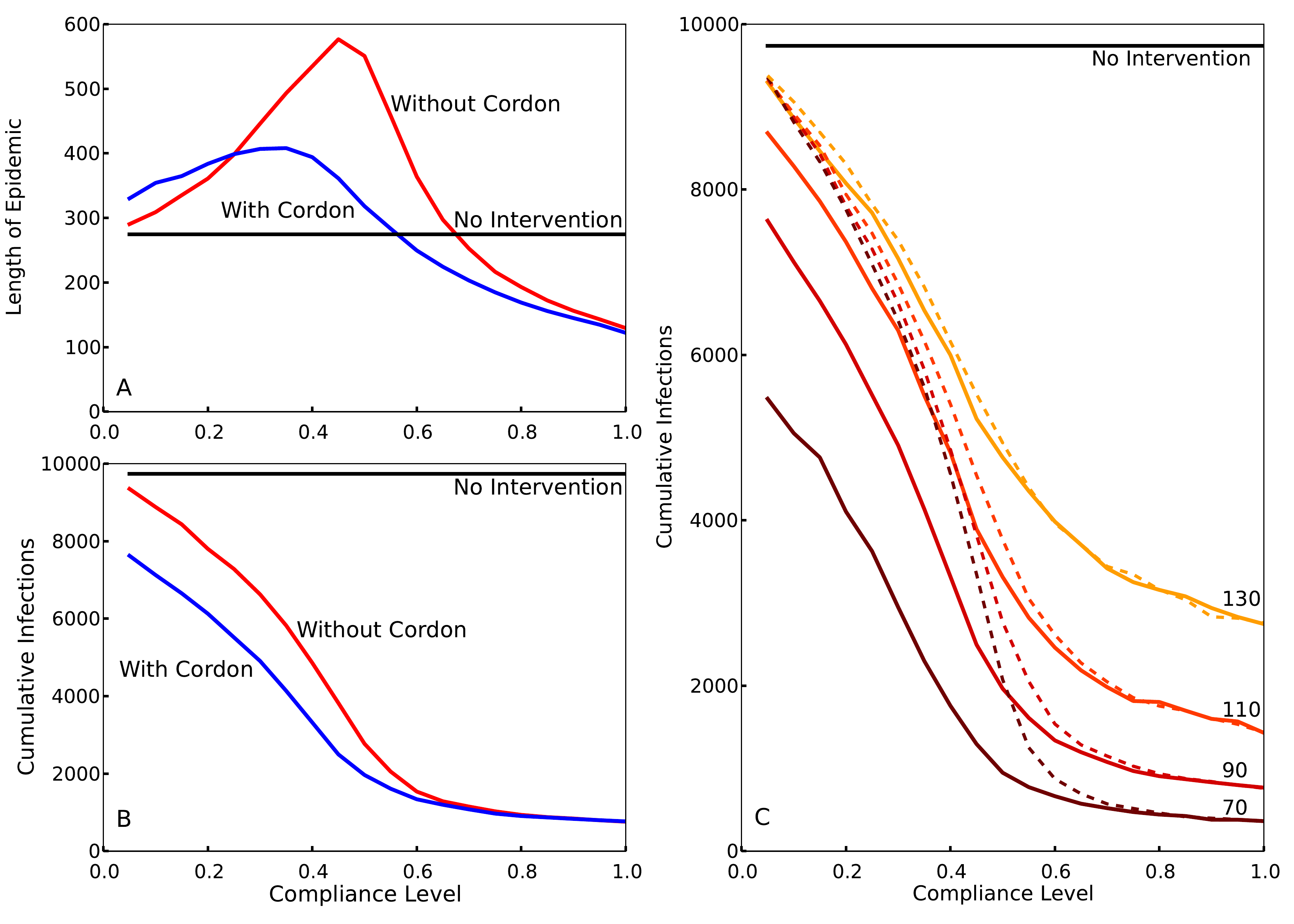}
\caption{\label{fig:3appendix} \textbf{Effect of compliance on epidemic length and cumulative infections with and without travel restrictions for the second set of paramter values ($\Delta = 10$, $\Gamma = 7$).} A,B. Simulations with (blue) and without (red) travel restrictions. The travel restrictions compensate for low levels of compliance, and their differences are comparable to Fig. \ref{fig:3} in the main paper. C. The cumulative number of infections over the entire epidemic, as a function of compliance levels and intervention times. Colors from brown to yellow signify intervention times (70, 90, 110, 130). Without enforced travel restrictions (dotted lines), a low compliance results in minimal differences between early and late policy implementation. Travel restrictions (solid lines) dramatically reduce infection numbers for earlier interventions at low compliance. We chose a slightly later set of intervention times $T_0$ for this set of parameters because the mean generation length ($\Delta + \Gamma$) \cite{chowell2014transmission, PMID:17476782} is about 50\% longer, 17 days, compared to the 11 days for the $\Delta = 5$, $\Gamma = 6$ case, so the exponential growth phase begins at a later time. }
\end{center}
\end{figure}

We simulated our epidemic with the parameters values $\Delta = 10$, $\Gamma = 7$, $\tau = 0.18$, and $\eta = 0.015$. 
As shown in Figs. \ref{fig:1appendix} and \ref{fig:3appendix}, the qualitative behavior of the epidemic and the impact of intervention policies were similar to the $\Delta = 5$, $\Gamma = 6$ case. For compliances of 0.6, 0.8, and 1.0, the outbreak ended quickly after the implementation of community-level isolation policies (Fig. \ref{fig:1}A), and the value of $R_t$ dropped well below 1 after a few days. The primary difference between this case and the one reported in the main paper can be seen in Fig. \ref{fig:1appendix}C where $R_t$ remains near 1 for a long period of time for 0.4 compliance (compare Fig. \ref{fig:1appendix}A). 

From Fig. \ref{fig:3appendix}C, we see that a compliance of 0.6 or higher limits the number of cumulative infections, and that travel restrictions still substantially help to limit the loss of life in the case of low compliance. Early implementation of the intervention (lower values of $T_0$) results in lower infection totals. However, we also see that cumulative case counts without travel restrictions (dotted lines) only begin to coincide with those with travel restrictions (solid lines) at higher levels of compliance than in the $\Delta = 5$, $\Gamma = 6$ case. This implies that the travel restrictions provide benefit at higher levels of compliance for these parameter values. Since these parameter values reflect a higher $R_0$ value (2.04) than the $\Delta = 5$, $\Gamma = 6$ case, this suggests that travel restrictions are more critical to halting outbreaks of more virulant diseases, as is to be expected. 

\bibliographystyle{Science}
\bibliography{Blah}

\end{document}